\documentclass[prb,twocolumn,showpacs,floatfix,amsfonts]{revtex4}
\usepackage{graphicx}
\usepackage{epsfig}
\begin{document}
\title[Thermal Conductivity Tensor]{Thermal Conductivity Tensor in YBa$_2$Cu$_3$O$_{7-x}$: Effects
of a Planar Magnetic Field}
\author{Roberto Oca\~{n}a and Pablo Esquinazi}
\affiliation{Abteilung  Supraleitung und
Magnetismus, Institut f\"ur Experimentelle Physik II,
Universit\"at Leipzig, Linn{\'e}str. 5, D-04103 Leipzig,
Germany}

\begin{abstract}

We have measured the thermal conductivity tensor of a twinned
YBa$_2$Cu$_3$O$_{7-x}$ single crystal as a function of angle
$\theta$ between the magnetic field applied parallel to the
CuO$_2$ planes and the heat current direction, at different
magnetic fields and at T=13.8 K. Clear fourfold and twofold
variations in the field-angle dependence of $\kappa_{xx}$ and
$\kappa_{xy}$ were respectively recorded in accordance with the
$d$-wave pairing symmetry of the order parameter. The oscillation
amplitude of the transverse thermal conductivity $\kappa^0_{xy}$
was found to be larger than the longitudinal one $\kappa^0_{xx}$
in the range of magnetic field studied here ($0~T~$$ \le B \le
9~$$T$). From our data we obtain quantities that are free from non-electronic contributions and
they allow us a comparison of the experimental results with current models
for the quasiparticle transport in the mixed state.
\end{abstract}
\pacs {74.25.Fy,74.72.Bk,72.15.He}
\maketitle

\section{Introduction}

The processes which influence the thermal transport of the quasiparticles (QP)
under the application of a magnetic field and below the superconducting
critical temperature are a matter of current research in the physics of high-temperature
superconductors (HTS). One of the difficulties to obtain clear evidence for one or other
 mechanism from thermal conductivity measurements that involves the QP, is basically
related to the separation of the QP contribution from the measured thermal conductivity. Due to the
fact that the experiments measure the total thermal conductivity given by the
sum of the contributions due to phonons, $\kappa^{\rm ph}$, and to the
QP, $\kappa^{\rm el}$, special methods are necessary to separate
the relatively large phonon contribution to the thermal transport in HTS.
Basically two methods for this separation have been treated in the literature. One method is
based on a phenomenological description of the field dependence of the electronic 
contribution of the longitudinal thermal
conductivity $\kappa_{xx}^{\rm el}(T,B)$,
introduced  first by
Vinen et al. \cite{vinen} and used in Refs.~ \cite{pogo,yu,kris2} to estimate $\kappa_{xx}^{\rm el}(T)$.
The other method, presented by Zeini et al. \cite{zeini}, is
 based on simultaneous measurements of the longitudinal and transverse thermal
conductivity and the assumption of a field independent Hall relaxation time.

To study thermal transport properties which depend on the QP contribution we have
chosen  to measure the thermal conductivity tensor as a function of
angle $\theta$ between heat current direction and the magnetic field applied parallel to the CuO$_2$ planes
at different magnetic fields. Because the oscillation amplitude of the thermal conductivities as a function of angle is
related only to the QP contribution, most of our results are basically free from
 phonon contributions and can help to test scattering models
for the QP. We discuss our results below in terms of two mechanisms that influence the
behavior of QP in the
mixed state of HTS: Andreev scattering (AS) is
 currently discussed in the literature as possible
scattering mechanism of QP \cite{sal,yu} and the Doppler shift (DS) 
as the main effect that changes the energy spectrum
of the QP with field \cite{vol2,kub1,kub2}.

The influence of the scattering mechanism on the thermal properties depends
also on the symmetry of the order parameter.
In the last years a large number of measurements has confirmed
the $d-$wave pairing symmetry for most of the HTS.  
 Due to  the
$d-$wave symmetry of the superconducting order parameter in HTS,
extended quasiparticle states in the mixed state - associated 
with a gapless structure at the nodes -
 are expected to contribute
to the thermal transport.
 These measurements have been described in terms of a
DS in the energy spectrum of QP and the 
scattering of QP by vortices (AS) taking into account the
presence of line nodes on the Fermi surface, i.e., within a pure
$d_{x^2-y^2}$-pairing symmetry \cite{chi,aub,yu}.

Try-crystal phase-sensitive measurements \cite{tsu}
have determined that the pairing symmetry of the order parameter is
$d_{x^2-y^2}$, which predominates with a small, if any, imaginary
component ($<5\%$) $id_{xy}$ at $T<T_c$.
Although symmetry experiments should not be
significantly affected by the development of a small $i d_{xy}$
contribution, the quasiparticle transport might
show (specially at $T \ll T_c$) quantitative differences from the
single $d_{x^2-y^2}$-pairing state due to the absence of line nodes
in the resulting superconducting $d_{x^2-y^2} + i d_{xy}$-gap.
This new state proposed originally by Laughlin \cite{lau} could
arise either by decreasing temperature or increasing magnetic
field as argued in Refs.~\cite{kas,voj1,voj2}.
 Up to now, however, there is not enough experimental evidence for such a
state although results on the magnetic field dependence of the
longitudinal thermal conductivity in BSCCO \cite{kris3,tal} and
temperature dependence of the Hall angle \cite{oca1} could be
explained taking into account the violation of both parity and
time-reversal symmetry. In this article we recover this idea and
attempt to establish the highest value, if any,  for a component
$id_{xy}$ of the superconducting order parameter within the
current models for the thermal transport in the HTS
cuprates.
 Our results agree with the
$d_{x^2-y^2}$-pairing symmetry of the order parameter and provide a
maximum value of $\sim35\%$ for the component $id_{xy}$.

From the phenomenological side our work provides interesting details. Namely, 
we show explicitly how the transverse
thermal conductivity $\kappa_{xy}$ can be calculated from the
experimental transverse temperature gradient $\nabla_y T$ and the
longitudinal thermal conductivity $\kappa_{xx}$ for magnetic
fields applied parallel to the CuO$_2$ planes. As in the case of a
perpendicular magnetic field, the
resulting $\kappa_{xy}$ depends on the ratio of the transverse
temperature gradient to the longitudinal one and  represents a physical property of the
 crystal involving only contributions from the
QP \cite{oca2}.
We also show
that the oscillation amplitude of the transverse thermal
conductivity, $\kappa^0_{xy}$, is larger than the longitudinal one,
$\kappa^0_{xx}$, in the whole magnetic field range studied
(0~T~$ \le B \le 9~$T).  From the extracted electronic contribution of
$\kappa_{xx}$ we attempt also to calculate a Hall-like angle and
show that  it increases
with magnetic field in qualitatively agreement with the AS model.

Our article is organized as follows. In the next section we describe some
experimental and sample details; in Sec.~\ref{res}  we describe and discuss
our results and is divided in four subsections. A summary is presented in
Sec.~\ref{sum}

\section{Experimental and Sample details}
\label{exp}

For all the measurements presented here we used a twinned
YBa$_2$Cu$_3$O$_{7-x}$ single crystal with dimensions (length $(l)
\times$  width $(w) \times$ thickness $(d)$) $0.83 \times 0.6
\times 0.045~$mm$^3$ and critical temperature $T_c =93.4~$K,
studied previously \cite{tal,oca1}. With polarized light microscopy we have determined
the positions of the twinning planes. Accordingly, the heat current direction was adjusted along
the $a/b$ axes which were parallel to the width or length of the sample.
The use of a twinned crystal rules out the influence of orthorhombicity on
the thermal transport properties studied in this work. In particular we can assume
that $\kappa_{xx}(T,B) = \kappa_{yy}(T,B)$. We note that the influence of the
orthorhombicity can be observed in untwinned crystals only if the impurity
scattering rate is small enough. As shown in Ref.~\cite{oca2}, untwinned
crystals with relatively large impurity scattering show similar angle dependences as
twinned ones.

\begin{figure}
\begin{center}
\centerline{\psfig{file=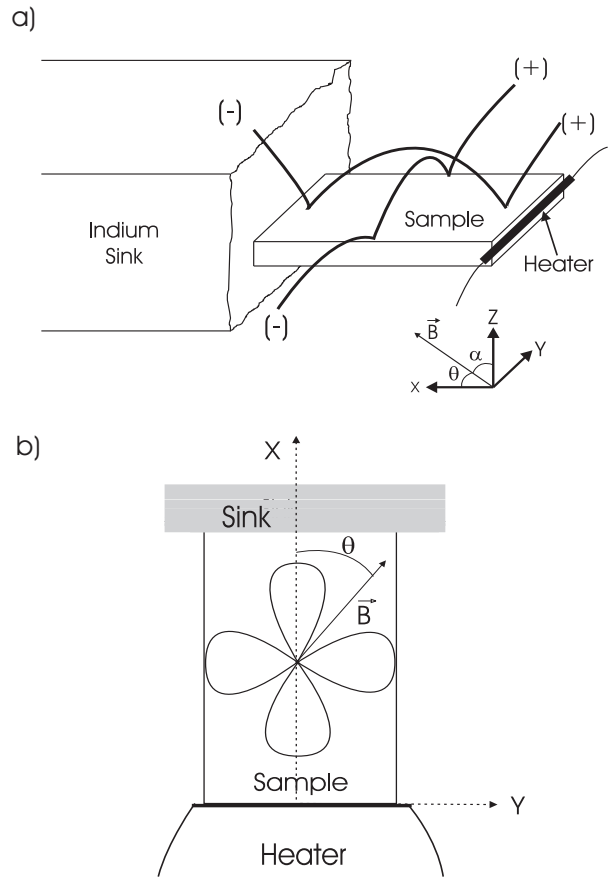,width=85mm}}
\end{center}
\caption[*]{(a) Experimental arrangement, sample axes and thermal
gradient definitions. The $``+"$ and ``-" signs of the thermocouples refer to
the hot and cold ends when the magnetic field is applied at $45^\circ$. The longitudinal 
gradient along $x$ is always negative. The misalignment angle $\alpha$ was minimized step-by-step
measuring the angle dependence of the longitudinal thermal gradient till reaching a satisfactory
symmetrical curve. (b) Top view of the sample arrangement with a sketch of the $d-$wave symmetry of the
superconducting order parameter.}
\label{arra}
\end{figure}

The longitudinal $(\nabla_x
T)$ and transverse $(\nabla_yT)$ temperature gradients have been
measured at $T = 13.8 K$ with previously calibrated
chromel-constantan (type E) thermocouples \cite{iny} and a dc
picovoltmeter. Our definition of sample axes, temperature
gradients and field direction is similar to that used in
Ref.~\cite{oca2}. If the heat current $\dot{Q}$ is along  $+\widehat{x}$
and a positive $90^\circ$ angle $\theta$ on the $(x,y)$-plane is along $+\widehat{y}$, then
there is a positive transverse thermal gradient for the magnetic
field $B$ at $+45^\circ$, see Fig.~\ref{arra}.

The magnetic field was applied perpendicular to the
c-axis. Special efforts were made in order to minimize the
misalignment of the plane of rotation with the CuO$_2$ planes. We
estimate the misalignment angle to be smaller than $\alpha \simeq
0.5^0$. An in-situ rotation system
enabled us the measurement of the thermal conductivity as a
function of the angle $\theta$ defined between the applied field
and the heat flow direction along $+ \widehat{x}$, see Fig.~\ref{arra}. 
We used an initial temperature
gradient $\nabla_x T \simeq 241~$K/m  ($\Delta_x T \simeq 200$~mK
for our crystal) at $\theta = 90^{\circ }$, i.e. with the
magnetic field perpendicular to the heat flow, and we recorded angle
scans and magnetic field dependence with a field-cooled
procedure to avoid pinning effects \cite{tal,aub}.
That means that  the angle
and magnetic field scans at constant temperature were done in such a way
 that after taking a reading the sample was driven to its
normal state by heating up to a few Kelvin above the
superconducting critical temperature and immediately was cooled
down again  to 13.8 K at constant field.
We will show
in Sec.~\ref{pin} experimental evidence that demonstrates how
pinning of vortices influences the field dependence of the measured
oscillations, even for the case of parallel field used in this work.

The possible influence of demagnetization and flux pinning effects associated with
the rectangular shape of the sample, see the comment and reply in Ref.~\cite{yu}, has been
checked by Aubin et al.~\cite{aub}. The authors concluded that the relative orientation of vortices
and the heat current, and not the sample geometry, governs the angle dependence. In our case,
due to the relative large applied fields in comparison with the lower critical field $H_{c1}$ and the field-cooled
procedure, an influence of the magnetization and flux pinning on the angle dependences 
can be excluded.

\section{Results and Discussion}\label{res}
\subsection{Angle dependence of the longitudinal $\kappa_{xx}$ and transverse $\kappa_{xy}$
thermal conductivities}\label{ang}

\begin{figure}
\begin{center}
\epsfig{file=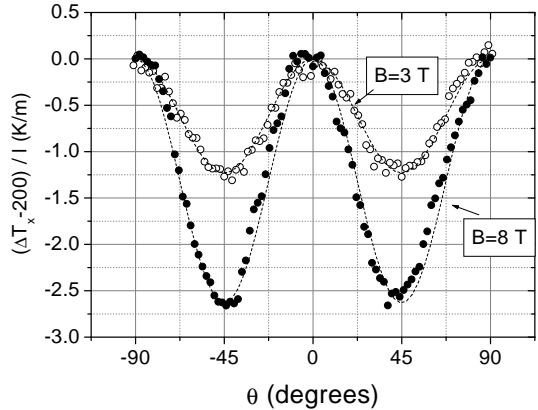,width=\columnwidth}
\end{center}
\caption[*]{(a) Field-angle dependence  of  the normalized temperature
gradient $(\Delta_x T-200)/l$ at 13.8 K for a magnetic field
strength of 8 T (solid circles) and 3 T (open circles). The heat
flow was constant during the whole scan and chosen to produce 200~mK
of temperature difference at the initial-scan angle $\theta =
90^{\circ }$. Dashed lines are fits to the function
$a (\cos(4\theta)-1)$ with a(T,B) as free parameter.}
\label{angle1}
\end{figure}

\begin{figure}
\vspace{-0.5cm}
\begin{center}
\epsfig{file=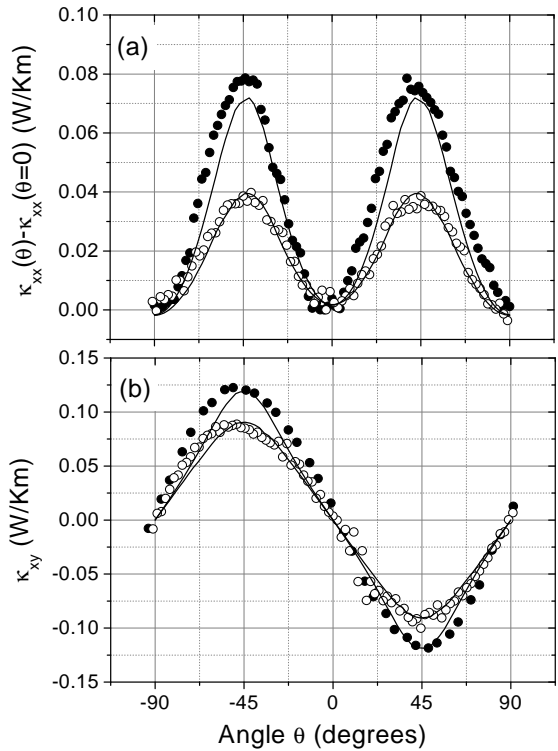,width=\columnwidth}
\end{center}
\vspace{-2.0cm}
\caption[*]{Angle dependence at two magnetic fields, $8$ T (solid circles) and $3$ T (open circles),
applied parallel to the
CuO$_2$ planes  of the (a) longitudinal and (b) transverse thermal conductivities.
The continuous curves are obtained from Eq.~(\protect\ref{brd}) with the parameters
given in the text and a pure $d_{x^2- y^2}-$wave order parameter.}
\label{kxxkxy}
\end{figure}

Figure \ref{angle1} shows the field-angle dependence of the
temperature gradient $(\Delta_x T-200)/l$ at 13.8 K for 8 T and
3T. As pointed out in Ref.~\cite{aub} the longitudinal thermal
conductivity $\kappa_{xx} \propto \dot{Q}/\Delta_x T$ and hence, the
longitudinal temperature gradient, shows fourfold symmetry with
minima (maxima for the thermal conductivity) along the nodal
directions of the superconducting gap, see Fig.~\ref{kxxkxy}(a).
We note that this effect represents a strong evidence for the
$d_{x^2-y^2}$-pairing symmetry of the order parameter, confirming
the presence of a large number of quasiparticles at those
particular orientations.

The curves in Fig.~\ref{angle1} can be fitted satisfactorily with the simple
function $a(\cos(4\theta)-1)$ with only $a$ as fit parameter which
depends on temperature and magnetic field. The fact that minima
are found in the field-angle profile of the temperature gradient
$(\Delta_x T-200)/l$ at the nodal directions could be explained
in terms of Andreev scattering of QP by vortices.
According to this effect if at any point the quasiparticle energy,
as viewed from the superfluid frame, equals the gap then, at that
point, Andreev reflection occurs: the quasiparticle is converted to
a quasihole reversing its velocity and its contribution to the
heat transport \cite{andi}. This mechanism of thermal resistance
is always induced when spatial variations of the amplitude or
phase of the order parameter take place. Since the QP
in the mixed state are in the presence of a phase gradient
(superfluid flow) they should also have a DS in their energy
spectrum given by the scalar product of the momentum and superfluid
velocity, $\bf p \cdot v_s$ as viewed from the laboratory frame.
That phase gradient is then the necessary mechanism that induces
the Andreev reflection. Thus, when a quasiparticle with momentum
{\bf p} is moving parallel to the magnetic field no DS
occurs and hence, no Andreev reflection. In other words, the field
acts as a filter \cite{mat} for that quasiparticle that contributes to
reduce the total temperature gradient. For different orientations
of the quasiparticle momentum with the field, there is a finite
probability for the Andreev reflection and thereby, for an enhancement of
the thermal resistance. The directionality of this scattering
mechanism is therefore the responsible feature for the
identification of the preferential directions of the
QP  by rotating the
magnetic field parallel to the basal planes.

On the other hand, a  DS in the energy spectrum of the
quasiparticle could also give rise to fourfold symmetry in the
longitudinal thermal conductivity by rotating the field parallel
to the CuO$_2$ planes, but in principle, with opposite sign. The
usual effect would be to produce an excess of quasiparticles in
the direction perpendicular to the field, and thereby, reducing
the thermal resistance in that direction \cite{won,hir1}. However,
the DS affects both the carrier density as well as the  scattering
rate of  QP, and since the latter dominates at high enough
temperatures the QP move easier parallel to the field and the sign
turns out to the correct one \cite{won2,hir1}. We note that,
although it is not clear which effect (AS or DS) should prevail at
low fields, the quasiparticle mean free path shortens with
increasing field as might be expected if the vortex are moving
closer together and therefore, AS should dominate the heat
transport at high enough fields.

\begin{figure}
\begin{center}
\epsfig{file=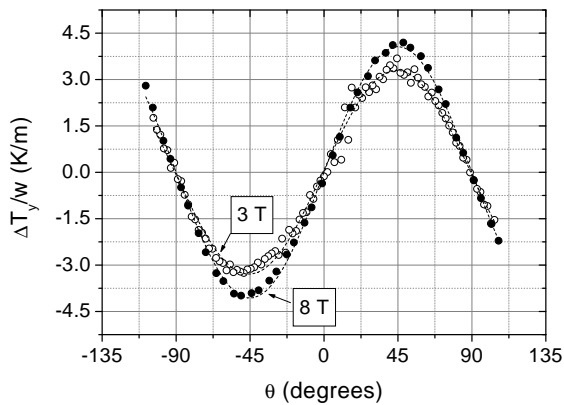,width=\columnwidth}
\end{center}
\caption[*]{Field-angle dependence of the transverse temperature
gradient $\Delta_yT/w$ at 13.8 K for a magnetic field strength of
$8$ T (solid circles) and $3$ T (open circles). As in Fig.~\protect\ref{angle1},
the longitudinal temperature difference at the beginning of the scan (at $\theta = 90^o$) 
was 200~mK. Dashed
lines are fits using $b \sin(2\theta)$ with b(T,B) as free
parameter.}
\label{angle2}
\end{figure}

In Fig.~\ref{angle2} we show the field-angle dependence of the
transverse temperature gradient $\Delta_yT/w$ at two magnetic fields.
We note that the sign of this effect is
just that expected within the pure AS picture,
i.e., the QP with momentum parallel to the magnetic
field does not suffer Andreev reflection, and therefore, enhances
the transverse temperature gradient in that direction. In other
words, the temperature of a side of the sample is enhanced when
the field is oriented to that side. As argued in Ref.~\cite{oca2}
and since
the transverse signal is expected to be free of non-electronic
contributions in contrast to the longitudinal temperature
gradient, this fact could be definitive in
distinguishing between Andreev reflection and DS. However,
 the inclusion of an impurity scattering rate in the
calculations of the DS effect accounts for the sign of both
fourfold and twofold oscillations as well
\cite{kub1,hir1,won,won2}. Furthermore, in the case of BSCCO it
has been argued that at the range of magnetic field and
temperature studied here, both effects are taking place, and
depending on the values of the parameters governing the gap at the
nodes, i.e, $v_F/v_\parallel$ and $k_F$, along with an impurity scattering
rate $\Gamma_0$,  different regimes in which DS dominates over
Andreev scattering or vice versa, could be identified \cite{vek2}.
Unfortunately, there is no yet theory for both longitudinal and
transverse thermal conductivities which takes into account both
effects when the field is oriented parallel to the CuO$_2$ planes.
Also, the theory for the thermal transport
for our case (i.e. parallel field, low temperatures and finite impurity scattering) 
under the influence of DS is still not yet developed. Therefore, in most of this paper we
restrict ourselves to a quantitative comparison of our experimental results to the
predictions obtained from the AS mechanism; deviations
from the predictions may well indicate that the other mechanism
could influence the thermal transport, as suggested in
Ref.~\cite{oca2}.

For a better comparison with theoretical predictions we need to
obtain the transverse thermal conductivity from the measured
transverse temperature gradient. As in the case the field is
applied normal to the basal planes, we calculate the transverse
thermal conductivity, $\kappa_{xy}$ as a function of the field
angle $\theta$ from the Onsager relations, using the fact that the
net heat current flowing in the $y$-direction is zero. At constant
temperature, this can be written as follows:
\begin{eqnarray}
\kappa_{xy}(B,\theta) = \kappa_{xx}(B,\theta) \frac{\nabla_y
T(B,\theta)}{\nabla_x T(B,\theta)}
\simeq \nonumber \\
 -\frac{J \nabla_y T(B,\theta)}{(\nabla_x T(B))^2}\,, \label{kxy}
\end{eqnarray}
where we have used the experimental fact that the total
longitudinal temperature gradient ($\sim 241$ K/m) is much larger
than its fourfold oscillation ($\sim 3$ K/m) as shown in Fig.~\ref{angle1}.
This allows us to approximate $(\nabla_x T(B,\theta))^2
\simeq (\nabla_x T(B))^2$ in (\ref{kxy}) and thereby, the fourfold
symmetry does not enter when calculating $\kappa_{xy}$. In other
words, the transverse thermal conductivity has the same symmetry
as the transverse temperature gradient but opposite sign. We note
that (\ref{kxy}) ensures us that $\kappa_{xy}$ measures in fact
a physical properties of the crystal as in the usual Righi-Leduc
effect. Figure~\ref{kxxkxy}(b) shows the angle dependence of  the transverse thermal conductivity
at two different magnetic fields.

\subsection{Separation of the phonon and QP contributions `a la Vinen' and comparison
with the Andreev-scattering mechanism}
\label{sep}

\begin{figure}
\begin{center}
\epsfig{file=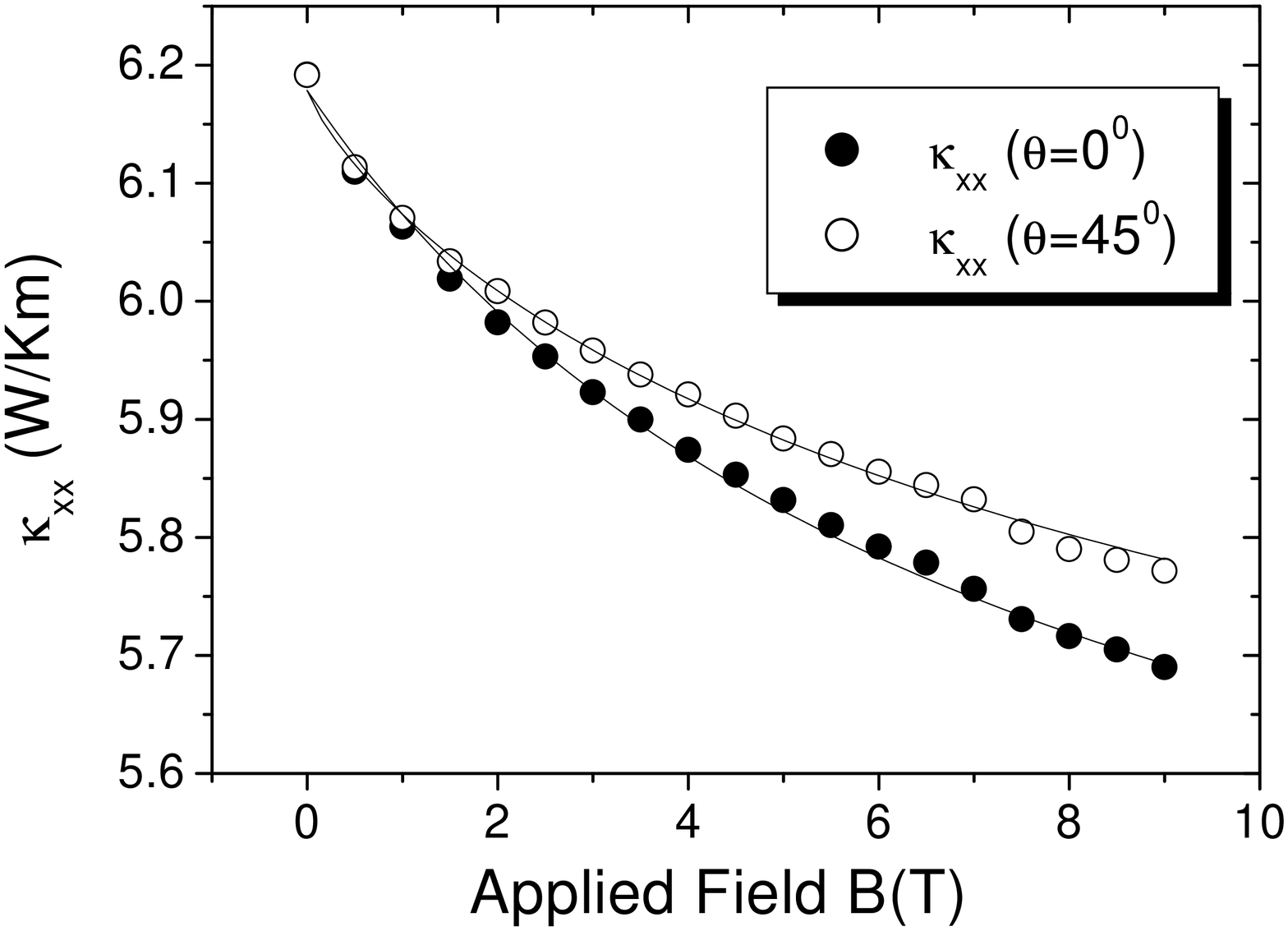,width=\columnwidth}
\end{center}
\caption[*]{Magnetic field dependence of the longitudinal thermal
conductivity $\kappa_{xx}$ with the field oriented along the
direction of the heat current ($\theta = 0^{\circ }$) and along a
nodal direction ($\theta = 45^{\circ }$). The two continuous
curves are obtained with (2) with $n = 1 (0.82)$ for the angle $\theta = 0^{\circ} (45^{\circ})$.}
\label{bdepen}
\end{figure}

In order to separate the phonon and quasiparticle contributions from
the longitudinal thermal conductivity, needed for a quantitative comparison with
the AS model, we have measured the field
dependence up to 9 T with the field oriented along both antinodes
and nodes directions. As shown in Fig.~\ref{bdepen} the field
profile can be rather well fitted with the empirical expression \cite{vinen}
\begin{equation}
\kappa_{xx}(T,B) = \kappa_{xx}^{\rm ph}(T) +
\frac{\kappa_{xx}^{\rm el}(T)}{1 + \beta_e(T) B^n}\,, \label{kapa}
\end{equation}
where, in general, $\beta_e(T)$ is proportional to the zero-field
electronic mean-free-path of the QP and $n$ can be
related to the nature of the quasiparticle scattering
\cite{pogo,vinen,yu}.

It is clear  that the phononic and
electronic contributions {\em at zero field}, $\kappa_{xx}^{\rm ph}$ and
$\kappa_{xx}^{\rm el}$, are independent of the angle $\theta$. Further, we assume that
 $\beta_e(T)$
remains unchanged by rotating the field parallel to the CuO$_2$
planes at constant temperature. In this case the exponent $n$
might be field-angle dependent due to the different types of  {\em
effective} scattering which the QP experience at different
orientations of the magnetic field. This assumption remains
thereby valid either when the main effect is the DS or the AS. As
in Ref.~\cite{yu} we used the value $n = 1$, when the field was
aligned parallel to the direction of the measured temperature
gradient (parallel to the heat current,
$\theta = 0^{\circ}$) and it was allowed to take other values for
different orientations of the field. As shown in
Fig.~\ref{bdepen}, we get satisfactory fits to the experimental
points with $\kappa_{xx}^{\rm ph} = 5.29$ W/Km,
$\kappa_{xx}^{\rm el} = 0.89 $W/Km, and $n \simeq 0.82$, the
latter when the field is oriented along the nodal direction
$\theta = 45^{\circ }$. To the best of our knowledge, it appears
that this is the first time such a dependence of the exponent $n$
in (\ref{kapa}) with angle $\theta$ is reported.

The microscopic model
for the scattering rate of the QP by vortices using Andreev reflection $\Gamma_{v}
(B,{\bf p})$ was formulated by Yu {\it et al. }\cite{yu}. The thermal conductivity 
can be calculated using a $2$D version
of the Bardeen-Richayzen-Tewordt model \cite{bar,yu},
\begin{equation}
\kappa^{el}_{\alpha \beta} = \frac{1}{2\pi^2ck_BT^2\hbar^2}
\int^\infty_{p_F} {\rm d}^2p \frac{v_{g\alpha}v_{g\beta} E_{\bf
p}^2}{\Gamma({\bf B},{\bf p},T)} {\rm sech}^2 \left(\frac{E_{\bf
p}}{2 k_B T} \right) \,, \label{brd}
\end{equation}
where $\Gamma ({\bf B},{\bf p},T)$ may be taken as a relaxation rate given by the sum of the
following scattering mechanisms acting in series: scattering of
QP by impurities $\Gamma_{\rm imp}({\bf p})$, by phonons
$\Gamma_{\rm ph}({\bf p},T)$, by QP $\Gamma_{\rm qp}({\bf
B},{\bf p},T)$ and AS by vortex supercurrents
$\Gamma_{v}({\bf B},{\bf p},T)$ \cite{yu}. The parameter $E_{\bf p}$ is
the QP energy,  $\alpha$ and $\beta$
denote the $x$ or $y$ directions on the plane of the sample,
$v_{g\alpha}$ is the group velocity along the $\alpha$ direction
and c is the c-axis lattice parameter. In Ref.~\cite{yu} an approximate
expression for the effective rate of AS by vortex current
was obtained which is basically given by
\begin{equation}
\Gamma_v \propto a_v^{-1} \exp(-a_v^2 f(E_{\bf p}), \phi) / \ln(a_v/\xi))\,.
\label{gam}
\end{equation}
In this expression the effective flux-line-lattice parameter is given by $a_v^2 \propto B_{c2}/B \gamma$,
where $B_{c2}$ is the upper critical field parallel to the planes and an effective mass factor
$\gamma \sim (m^*/m)^2$.
The function $f$ depends on the energy of the QP, and consequently on the 
gap $\Delta({\bf p})$, and the angle $\phi$ between the direction of
the momentum of the QP and the magnetic field. For more details see Ref.~\cite{yu}.
As we mentioned above, we would like to know how large, if any, can be the contribution 
of a term $id_{xy}$ in the order parameter compatible with our transport measurements.
Therefore we use the following gap
\begin{equation}
\Delta({\bf p},T)\,=\,\Delta(T)(\Delta_{d_{x^2-y^2}}+i\Delta_{d_{xy}})\,,
\end{equation}
which can be explicitly written taking into account the momentum
components as
\begin{widetext}
\begin{equation}
\Delta({\bf p},T)\,=\,\Delta(T)\left (\frac{\cos({\frac
{p_xa}{\hbar}})-\cos({\frac {p_ya}{\hbar}})}{1-\cos( {\frac
{p_F\,a}{\hbar}})}+i\,\frac{2\sin(\,{\frac {p_xa}{2\hbar}})\sin({
\frac {p_ya}{2\hbar}})}{1-\cos({\frac {p_F\,a}{\hbar}})}\right )\,.
\label{gap}
\end{equation}
The percentage values indicated below for the $id_{xy}$ part
means its weighted contribution relative to the total gap. 
\end{widetext}

Our measurements are done at constant temperature, therefore if we make
the simplifying assumption that the scattering rates
$\Gamma_{\rm imp}$, $\Gamma_{\rm ph}$ and $\Gamma_{\rm qp}$ do not depend
on the momentum of the excitations, we can write
$\Gamma_{\rm imp}+\Gamma_{\rm ph}+\Gamma_{\rm qp} \equiv \Gamma_0 \simeq cte$.
At higher temperatures it would be necessary,  however, to include a
model for the energy dependence of the inelastic scattering rate
$\Gamma_{\rm qp}$ as argued by many authors \cite{yu2,hir2}.
Nowadays there is consent that
the peak observed in the longitudinal thermal conductivity as a
function of the temperature is due to the  competition between
a decreasing inelastic scattering rate and the decrease in the
number of quasiparticles with decreasing temperature.

From (\ref{brd}) we obtain $\kappa_{xx}^{\rm el}\Gamma_0 = 7.2
\times 10^{12} $(W/Kms) at zero field assuming
$d_{x^2-y^2}$-pairing symmetry and maximum gap $\Delta(T = 0) = \Delta_0 = 20$ meV.
Then, if we use the quasiparticle contribution $\kappa_{xx}^{\rm el}$
obtained from the fits using (\ref{kapa}), we get $\Gamma^{-1}_0
\simeq 0.12$~ps. Using this value and the parameters from
Ref.~\cite{yu}, i.e. $B_{c2} = 500$~T, Ginzburg-Landau parameter
$\kappa = 100$, $\gamma = 4$ and $\Gamma^{-1}_{v}(B = B_{c2})
\simeq 3/2\Gamma^{-1}_0$ (chosen to reproduce the experimental amplitude of
the oscillation at 8 T) it is not possible to fit satisfactorily
the thermal conductivity $\kappa_{xx}(B,\theta)$ of
Fig.~\ref{kxxkxy}(a) for both curves at 3T and 8T simultaneously.
However, if we increase strongly the upper critical field to
an unrealistically high $B_{c2}  \simeq 9,500~$T or increase $a_v$ by a factor $\sim 5$,
Eq.~(\ref{brd}) accounts for the fourfold
oscillation observed in $\kappa_{xx}(B)$ (and hence, for the
normalized temperature gradient $(\Delta_x T-200)/l$)) when the
field is rotated from $\theta = +90^{\circ }$ to $\theta =
-90^{\circ }$ as shown in Fig. \ref{kxxkxy}(a).

With the same parameters mentioned above, $\kappa_{xy}$
can be calculated from (\ref{brd}) as shown in the Fig.~\ref{kxxkxy}(b).
We find agreement
between the experiment and theory if, as for $\kappa_{xx}$,  we increase the
intervortex spacing $\sim 5$ times the value defined in Ref.~\cite{yu,aub}. This
result indicates that the AS rate needed to fit the experimental
data is much smaller than the one obtained from the  proposed microscopic model
and with the parameters used in Ref.~\cite{yu}.

\begin{figure}
\begin{center}
\epsfig{file=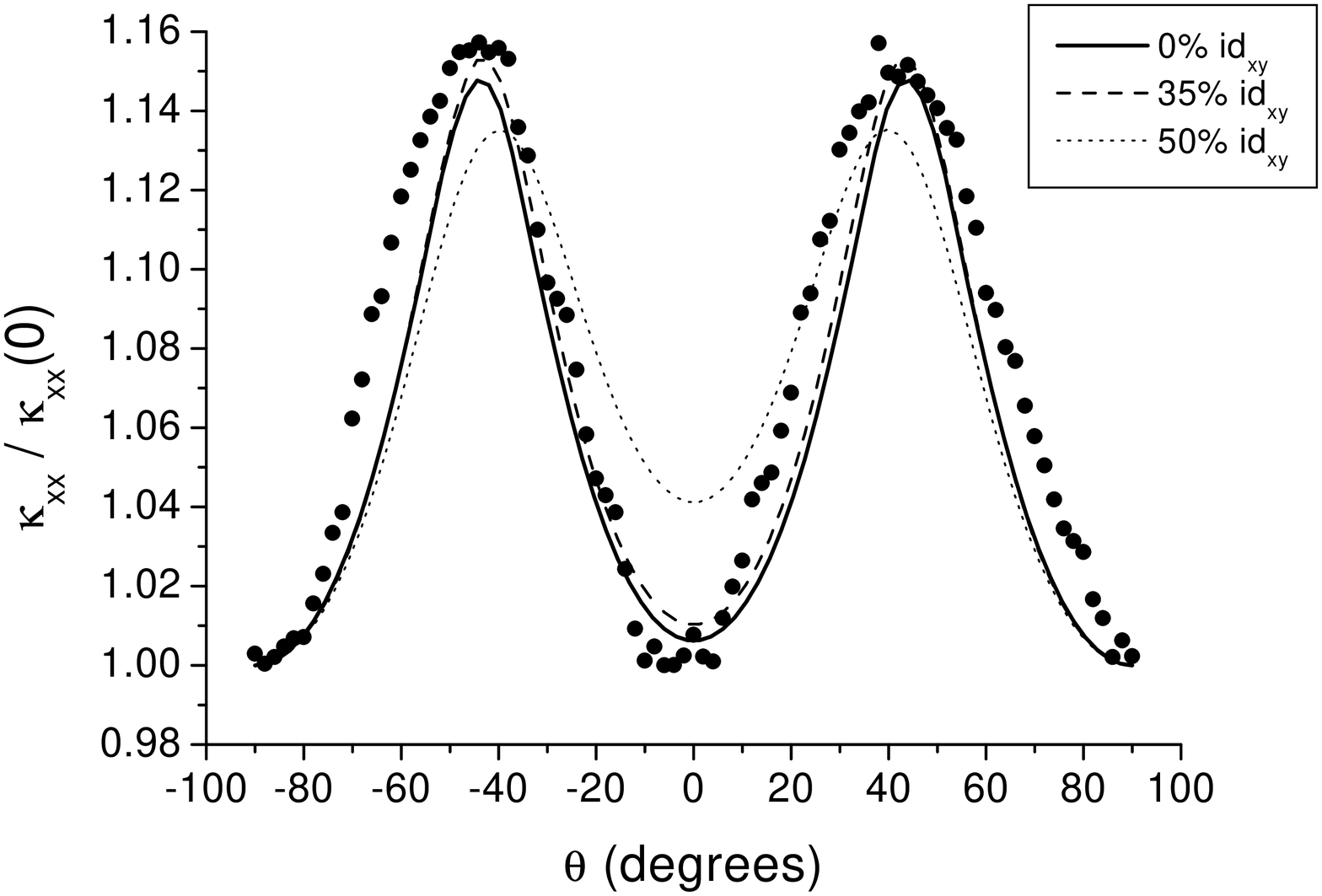,width=\columnwidth}
\end{center}
\caption{Angle dependence of the normalized longitudinal conductivity at a field of 8~T
$\kappa_{xx}(\theta)/\kappa_{xx}(0)$. The data points are taken from Fig.~\protect\ref{kxxkxy}(a)
normalized by an arbitrary factor in order to compare with the theoretical predictions given
by the continuous, dashed and dotted lines. These lines were calculated with (\protect\ref{brd})
assuming a pure $d_{x^2-y^2}$ symmetry, plus a 35\% and $50\%$~$id_{xy}$ contribution added 
to the gap (see Eq.~(\protect\ref{gap})),
respectively, and with the same parameters used to fit the data of Fig.~\protect\ref{bdep}.} \label{idxy}
\end{figure}

We can also extend the AS model  including in (\ref{brd}) an 
 additional component $i d_{xy}$ to the pure $d_{x^2-y^2}$-gap and compare it
 to the experimental angle dependence of  $\kappa_{xx}(\theta)$, for example. 
The results shown in Fig.~\ref{idxy}  indicate 
that the fourfold oscillation is no longer observable with a
component larger than $\sim 35 \%$  at 13.8 K. Because we compare the symmetry of 
both theoretical and experimental curves 
taking the normalized $\kappa_{xx}$ (see Fig.~\ref{idxy}), this upper limit
is not affected using other parameters in Eq.~(\ref{brd}).

\subsection{Field dependence of the oscillation amplitudes $\kappa^0_{xx}$
and $\kappa^0_{xy}$:  Doppler shift and the effect of pinning of vortices}
\label{pin}

\begin{figure}
\begin{center}
\epsfig{file=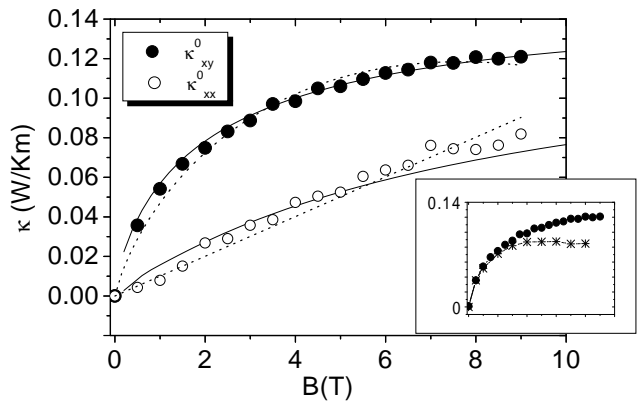,width=\columnwidth}
\end{center}
\caption{Magnetic field dependence of
$\kappa^0_{xx}$ (open circles) and $\kappa^0_{xy}$ (solid
circles). Solid lines are fits using Eq. (\ref{brd}). 
Dashed lines are fits using the DS model following Eqs.~(\protect\ref{ds1}).
The inset
shows the amplitude $\kappa^0_{xy}$ recorded with a non-field
cooled procedure ($*$) and with a field-cooled procedure
(solid circles, the same data as in the main figure).} \label{bdep}
\end{figure}

Other way to compare theory and
experiment that does not need any empirical relation to separate the QP contribution
is based on the measurement of the field strength
dependence of both fourfold and twofold oscillation amplitudes, which are in
fact pure electronic effects. It is convenient to
define the quantities
\begin{eqnarray}
\kappa^0_{xy} \equiv \kappa_{xy}(\theta =
0^{\circ }) -  \kappa_{xy} (\theta = 45^{\circ })\,,
\nonumber \\
\kappa^0_{xx} \equiv \kappa_{xx} (\theta = 45^{\circ }) -
\kappa_{xx} (\theta = 0^{\circ })\,,
\label{def}
\end{eqnarray}
 which can be directly compared
with any  model that accounts for the contribution  of the QP. In
Fig.~\ref{bdep} we show the magnetic field dependence of
$\kappa^0_{xy}$ and $\kappa^0_{xx}$. As shown, we find that the
resolution of $\kappa^0_{xx}$ is smaller than that of
$\kappa^0_{xy}$. The fourfold oscillation observed in
Fig.~\ref{angle1} has a thermal background of $\sim 200 $ mK,
which makes more difficult the observation of any small superposed
signal. The thermal background in the transverse thermocouple due
to a small misalignment in the $x$-direction was $\sim 8.5 $mK.

We note that the oscillation amplitude of the transverse thermal
conductivity $\kappa^0_{xy}$ is larger than the one  of the longitudinal thermal conductivity
$\kappa^0_{xx}$.
Note the fact, that Eq.~(\ref{brd}) with the parameters
described above (including an intervortex spacing $\sim 5$ times larger) accounts
 for this behavior, see Fig.~\ref{bdep}. One would tend to believe that 
changing the effective mass factor $\gamma$ we may compensate the too high
$B_{c2}$-value needed to fit the experimental data in Fig.~\ref{bdep}. Note 
that $\gamma$ enters not only in $a_v$ but in other parts of Eqs.~(\ref{brd}) and (\ref{gam}).
Our numerical results show no agreement with the experimental data of Fig.~\ref{bdep} when we choose
$\gamma > 4$ changing $B_{c2}$ between 10,500~T and 500~T.

 With the proposed DS mechanism by Won and Maki \cite{won2} that would influence the quasiparticle 
transport we can
 also get satisfactory fits to our experimental points either for
 $\kappa_{xy}^0$ or $\kappa_{xx}^0$ but  not for both
 components with exactly the same parameters. 
As we noted above and in Ref.~\cite{oca2}, the predictions of the
DS model for the thermal conductivities with a parallel field were
obtained in the superclean limit (this means that no impurity scattering rate
appears explicitly in the thermal conductivity tensor) and for the
temperature range $k_B T \gg \epsilon = (1/2)\sqrt{vv'eB}$ ($v$ and $v'$ are the
Fermi velocities perpendicular and parallel to the Fermi surface, respectively) \cite{won2}: 
\begin{widetext}
\begin{eqnarray}
\frac{\kappa_{xy}}{\kappa_n} = -\frac{1}{(2\pi)^2}\frac{v v' e
H}{\Delta^2}\ln{(\frac{2\Delta}{1.75T})}\ln{(\frac{4\Delta}{v v'
\sqrt{eH}})}\sin{2\theta} \ , \nonumber
\end{eqnarray}
\begin{eqnarray}
\frac{\kappa_{xx}}{\kappa_n} =
\frac{7\pi^2}{10}(\frac{T}{\Delta})^2(1 +
(\frac{2}{\pi})^2\ln{(\frac{2\Delta}{1.75T})})  -
\frac{1}{2\pi^2}\ln{(\frac{2\Delta}{1.75T})}\frac{v v' e
H}{\Delta^2}[\ln{(\frac{4\Delta}{v v' \sqrt{eH}})} -
\frac{1}{16}(1 - \cos{(4\theta)})] \,, \label{ds1}
\end{eqnarray}
\end{widetext}
where $\Delta$ is the energy gap and, in units of $\hbar k_B$, $\kappa_n =  T n/3\Delta m$,   $n$ the density 
of QP and $m$ their mass. Both expressions provide the correct sign
for the tensor components. From these equations and as shown in Ref.~\cite{oca2} we obtain for the
field dependence of the oscillation amplitudes  $\kappa_{xy}^0 = 
a B \ln(1/bB)$ with $a$ and $b$ fit parameters, and 
 $\kappa_{xx}^0 = aB$. These dependences also agree well with the experimental
data, see Fig.~\ref{bdep}. The fit of
 $\kappa_{xy}^0$ gives $a \simeq 0.015$
 (W/TKm) and $b = 0.048 {\rm T}^{-1}$; from the expression proposed for $\kappa_{xx}^0$ we get $a \simeq 0.010$
 (W/TKm). This discrepancy in $a$ is not too large taking into account that neither
 the superclean limit used in developing this model nor the field-temperature range where the Eqs.~(\ref{ds1})
are valid,  reflect the situation studied here. Therefore, a more accurate
 treatment of the impurity scattering is necessary to check if the
 DS picture can fit quantitatively the experimental data. It is striking
 that the DS description appears also to provide the experimental field
 dependences of the oscillation amplitudes.

 In the
inset, we show the effect of the pinning of vortices on the oscillation
amplitude $\kappa^0_{xy}$. While the
points taken with a field-cooled procedure show a monotonic
behavior in the range of fields studied here, the points taken
with a non-field-cooled procedure, i.e. in this case the field 
was increased from zero at constant
temperature and always at the same initial angle,  show a saturation above
$\sim 3$ T. This effect could have a different origin from the
DS, and would indicate that vortex scattering
becomes more important with increasing field. We note that the pinning of
the Josephson-like vortices parallel to the planes is strongly affected
by vortices perpendicular to the planes due to the misalignment of the
crystal respect to the applied field. The results in the inset of Fig.~\ref{bdep}
clearly stress the necessity  of field-cooled experiments to get the true
magnetic field dependence of thermal transport parameters.

\subsection{Hall-like angle and quasiparticles-dependent ratios}

The angle dependent conductivities we have measured  allow us the definition of
quantities which just involve contributions from the QP
as in the usual Righi-Leduc
effect. As noted above we do not have to assume strictly any method to
separate electronic and phononic contributions from the total
thermal conductivity to obtain quantities which can be
directly compared with the theory of QP transport. As shown in the last section, some of these
quantities are themselves the field dependence of both
$\kappa^0_{xy}$ and $\kappa^0_{xx}$. However, the ratio
$\kappa^0_{xy}/\kappa^0_{xx}$ might provide also an interesting
result for comparing with theory. In fact, it describes an
intrinsic property of the crystal studied, free of the
experimental uncertainty to determine the total power given to the
sample.
\begin{figure}
\vspace{3cm}
\begin{center}
\epsfig{file=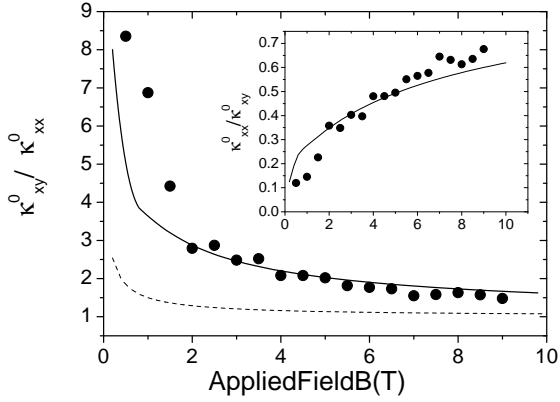,width=\columnwidth}
\end{center}
\vspace{-3.0cm}
\caption{Magnetic field dependence of the ratio
$\kappa^0_{xy}/\kappa^0_{xx}$ and $\kappa^0_{xx}/\kappa^0_{xy}$ (inset).
The continuous line is obtained with Eq.~(\protect\ref{brd}) and with the same
parameters of  Fig.~\protect\ref{bdep} described in the text, whereas the dashed line
is obtained with an upper critical field $B_{c2} = 650~$T.}
\label{ratio}
\end{figure}

As shown in Fig.~\ref{ratio}, two regimes in the oscillation ratio
$\kappa^0_{xy}/\kappa^0_{xx}$ can be distinguished: From zero to
$\sim 2$ T the ratio decreases stronger with field than at fields
above $\sim 2$ T. An abrupt decrease of this ratio is predicted by
Eq.~(\ref{brd}) but it fits well the experimental data only above
$\sim 2$ T, see Fig.~\ref{ratio}. This discrepancy at low fields
could indicate that at $B < 2~$T other mechanism influences the
QP transport and partially overwhelms the influence of the proposed AS.
In the same figure and to demonstrate the influence of the flux-line-lattice parameter on the
theoretical predictions we show the ratio (dashed line) calculated with the same parameters
as before but with $B_{c2} = 650~$T. 

In order to provide further analysis of our data, it is
interesting to define a Hall-like angle as in the case of a
normal field applied to the CuO$_2$ planes, and to explore its
magnetic field dependence. Therefore, since in this case the
angle $\theta$ is an intensive parameter of the measurement along
with the temperature $T$ and the magnetic field strength $B$, it
seems reasonable to perform an average on the angle dependence of both
conductivities over the range $[-90,90^{\circ
}]$ of $\theta$.
Also, we do not need to take into account
the sign of $\kappa_{xy}$ which indicates
 which side of the sample becomes hotter when the field is
applied, characterizing the deflection direction of the QP.  At constant temperature, we 
define the effective Hall-like angle as
\begin{equation}
\tan\alpha_{\parallel} \equiv
\frac{\langle\left|\kappa_{xy}(B)\right|\rangle_{\theta}}{\langle\kappa_{xx}^{\rm el}(B)\rangle_{\theta}}\,,
\label{alpha}
\end{equation}
where $\langle\ \rangle_{\theta}$ means the integration over $\theta$ in the
range mentioned above.

\begin{figure}
\begin{center}
\vspace{-2.0cm}
\epsfig{file=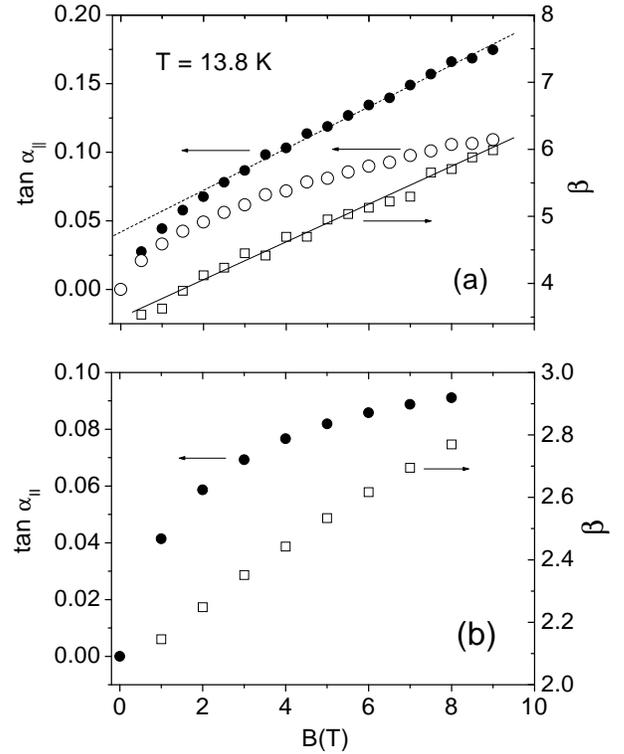,width=\columnwidth}
\end{center}
\vspace{-1.0cm}
\caption{(a) Left axis: Magnetic field dependence of
$\tan\alpha_{\parallel}$ obtained using the phonon thermal conductivity
$\kappa_{xx}^{\rm ph}$ obtained from the best fit of $\kappa_{xx}(B)$ using
(\protect\ref{kapa}) $(\bullet)$, or decreasing this by $5\%~(\circ)$.
Right axis: Magnetic field dependence of the ratio $\beta$. The dashed and straight lines
are guides to the eye. (b) The same as panel (a) but the points were obtained from Eq.~(\protect\ref{brd})
assuming pure $d_{x^2-y^2}$-wave symmetry and with the same parameters that fit the
field dependence of the oscillation amplitudes in Fig.~\protect\ref{bdep}.}
\label{tanal}
\end{figure}

In Fig.~\ref{tanal} we show the field dependence of
$\tan\alpha_{\parallel}$ in which we used the phononic
contribution to the thermal conductivity $\kappa_{xx}^{\rm ph}$
extracted using (\ref{kapa}) to calculate the total
quasiparticle contribution $\kappa_{xx}^{\rm el}(B,\theta)$. Note
 that the field dependence of $\tan\alpha_{\parallel}$ is
affected by  the value of $\kappa_{xx}^{\rm ph}$ obtained from (\ref{brd}), in particular
the linear dependence with the magnetic field observed above $\sim
2$ T. Decreasing 5\% the obtained $\kappa_{xx}^{\rm ph}$ the computed  
$\tan\alpha_{\parallel}$ decreases substantially at high fields, see Fig.~\ref{tanal}.
This result emphasizes the difficulty to obtain true field dependences 
of thermal transport parameters by using phenomenological separation
methods.

A magnitude free of  the separation model is given by
 the ratio between the averaged change with field of the
longitudinal thermal conductivity, $\langle\kappa_{xx}(B=0) -
\kappa_{xx}(B,\theta)\rangle_{\theta}$ and the averaged transverse
thermal conductivity
$\langle\left|\kappa_{xy}(B)\right|\rangle_{\theta}$, which only
contains information of the quasiparticle contributions. We define then,
\begin{eqnarray}
\beta \equiv \frac{\langle\kappa_{xx}(B=0) -
\kappa_{xx}(B,\theta)\rangle_{\theta}}{\langle\left|\kappa_{xy}(B)\right| \rangle_{\theta}} =
\nonumber \\
\frac{\pi}{2} \left [ \frac{\kappa_{xx}^{\rm el}(0,0) - \kappa_{xx}^{\rm el}(0,B)}{\kappa^0_{xy}} -
\frac{\kappa^0_{xx}(B)}{2 \kappa^0_{xy}(B)} \right ]
\,.
\label{beta}
\end{eqnarray}
As shown in Fig.~\ref{tanal} (right axis), the ratio $\beta$ increases
roughly proportional with field. Note first that the simple
subtraction $\langle\kappa_{xx}(B=0) -
\kappa_{xx}(B,\theta)\rangle_{\theta}$ cancels the phononic
contribution of the thermal conductivity which we assume it does
not change with field \cite{yu2,zeini,kris2,oca1}. Second, $\beta$ does not
provide the same information as the ratio $\kappa^0_{xx}/\kappa^0_{xy}$. Therefore,
$\beta$ can be used for comparing with any model for the
quasiparticle transport. 
We note that the observed
linear regime of $\beta$ should disappear with increasing field.
When the field is increased, the vortices are
moving closer together and hence the scattering probability 
of QP by vortices increases. Thus, at higher fields
than those used in the present work, either a saturation or a
non-monotonic behavior in the field dependence of $\kappa^0_{xy}$
as well as $\kappa^0_{xx}$ is expected. 

With Eq.~(\ref{brd}) and with the same parameters that fit the oscillation amplitudes (see Fig.~\ref{bdep})
we have calculated both parameters $\tan(\alpha_{||})$ and $\beta$, see Fig.~\ref{tanal}(b).
We found that the field dependence of both parameters is qualitatively reproduced by the
AS mechanism. However, both their absolute values and their slopes with field are not given by
Eq.~(\ref{brd}) correctly. A disagreement is also found if we want to fit
the curves given in Fig.~\ref{bdepen} using the same parameters as in Fig.~\ref{bdep}.
Clearly, the actual AS model is not enough to understand the experimental data quantitatively.

\section{Summary}
\label{sum} In summary, we have measured the components of the
thermal conductivity tensor at $13.8$ K in an optimally doped
twinned single crystal of YBCO as a function of the magnetic field
and the angle $\theta$ between magnetic field and heat current.
The field-angle dependence measurements show the predominance of
the $d_{x^2-y^2}$-wave pairing symmetry over a complex
component $id_{xy}$, which is estimated to be less than $\sim 35\%$ 
of the resulting total gap,
 within the AS picture for the interaction of  QP with vortices.

The magnetic field
dependence of the amplitude coefficients of the fourfold and
twofold oscillations observed in $\kappa_{xx}$ and $\kappa_{xy}$
respectively, allows us to define magnitudes free of contributions
from the phonons and, hence, suitable for comparison with theory.
In particular, the dependence of the 
amplitude coefficients $\kappa^0_{xy}$ and $\kappa^0_{xx}$, 
their ratio, a Hall-like angle $\tan\alpha_{\parallel}$ and the ratio $\beta$ between
the decrease of the longitudinal thermal conductivity and the
transverse thermal conductivity with  field were obtained.
Assuming
only the AS mechanism  along with an impurity scattering
rate, we  get a good quantitative
description of both field dependences of the oscillation amplitude of the conductivities
 only if  we increase the intervortex spacing
 five times (or increase the parallel critical field to 9500~T) the value defined in Ref.~\cite{yu}.
This implies  a much smaller scattering rate due to AS than the
one obtained from the original theory. 
Also, the AS model does not fit with the same parameters all the experimental data, although
it provides a good qualitative description of the field and angle dependences
of all the parameters.

The less developed  DS model does not allow a rigorous quantitative comparison.  It is worth
to mention, however, that the DS model alone also provides field dependences of the
oscillation amplitudes similar to those obtained experimentally. An impurity
scattering rate should be introduced in the DS theory within a more
realistic limit than the superclean limit to perform a
detailed discussion of the quantities involved in the equations. 
The overall results suggest that both AS
and DS effects are responsible for the observed magnetic field
dependencies  but with different regimes of predominance.
As suggested in
Ref.~\cite{vek2} it is more likely that the DS plays
an important role at low fields providing
a slower magnetic field dependence in the components of the
thermal conductivity tensor than the AS. This can
be understood taking into account that while the
QP always suffer a DS in the mixed state,
the probability for Andreev reflection increases with increasing
field. This might explain the less abrupt decrease with field of the ratio
$\kappa^0_{xy}/\kappa^0_{xx}$ below $\sim 2$ T and the 
observed quantitative discrepancies between the Hall-like angle
$\tan\alpha_{\parallel}$, the parameter $\beta$ and the AS model. Further development
of the theory  involving both mechanisms is necessary to confirm this hypothesis.

\begin{acknowledgments}
This work is supported by the DFG under Grant DFG Es 86/4-3 and
partially supported by the DAAD. We are specially grateful with I.
Maksimov and K. Maki for fruitful discussions. P.E. acknowledges the hospitality
of the Condensed Matter Physics Department (C-III) of the Universidad Aut\'onoma de Madrid
and the support given by the Secretar\'ia de Estado de Educaci\'on y Universidades 
(grant SAB2000-0139).

\end{acknowledgments}

\end{document}